\documentclass[12pt]{article}

\usepackage{graphicx}
\usepackage{subcaption}
\usepackage{dcolumn}
\usepackage{bm}
\usepackage{nicefrac}
\usepackage{color}
\usepackage{xcolor}
\PassOptionsToPackage{rgb}{xcolor}
\pdfpageattr {/Group << /S /Transparency /I true /CS /DeviceRGB>>}

\newenvironment{sciabstract}{%
	\begin{quote} \bf}
	{\end{quote}}

\usepackage[utf8]{inputenc}
\usepackage[T1]{fontenc}
\usepackage{mathptmx}

\begin{document}
\title{Novel beamline for attosecond transient reflection spectroscopy in a sequential two-foci geometry}

\author
{Giacinto D. Lucarelli$^{1,2}$, Bruno Moio$^{1,2}$, Giacomo Inzani$^{1}$,\\
	 Nicola Fabris$^{3}$, Liliana Moscardi$^{4}$, Fabio Frassetto$^{3}$, \\
	 Luca Poletto$^{3}$, Mauro Nisoli$^{1,2}$, Matteo Lucchini$^{1,2\ast}$\\
	\\
	\small{$^{1}$Department of Physics, Politecnico di Milano, 20133 Milano, Italy}\\
	\small{$^{2}$Institute for Photonics and Nanotechnologies, IFN-CNR, 20133 Milano, Italy}\\
	\small{$^{3}$Institute for Photonics and Nanotechnologies, IFN-CNR, 35131 Padova, Italy}\\
	\small{$^{4}$Center for Nano Science and Technology@PoliMi, Istituto Italiano di Tecnologia, 20133 Milano, Italy}\\
	\\
	\small{$^\ast$To whom correspondence should be addressed; E-mail:  matteo.lucchini@polimi.it.}
}

\date{\today}

\maketitle

\begin{sciabstract}
	We present an innovative beamline for extreme ultraviolet (XUV)-infrared (IR) pump-probe reflection spectroscopy in solids with attosecond temporal resolution. The setup uses an actively stabilized interferometer, where attosecond pulse trains or isolated attosecond pulses are produced by high-order harmonic generation in gases. After collinear recombination, the attosecond XUV pulses and the femtosecond IR pulses are focused twice in sequence by toroidal mirrors, giving two spatially separated interaction regions. In the first region, the combination of a gas target with a time-of-flight spectrometer allows for attosecond photoelectron spectroscopy experiments. In the second focal region, an XUV reflectometer is used for attosecond transient reflection spectroscopy (ATRS) experiments. Since the two measurements can be performed simultaneously, precise pump-probe delay calibration can be achieved, thus opening the possibility for a new class of attosecond experiments on solids. Successful operation of the beamline is demonstrated by the generation and characterization of isolated attosecond pulses, the measurement of the absolute reflectivity of SiO$_2$, and by performing simultaneous photoemission/ATRS in Ge.
\end{sciabstract}

\section{\label{sec:intro}Introduction}

In the last twenty years, the exploitation of high-order harmonic generation (HHG) in gases has opened new possibilities for the investigation of fundamental processes in atoms, molecules and solid-state materials with extreme temporal resolution \cite{Krausz2009,Nisoli2017}. High-order harmonics originate from the interaction of an intense ($\sim 10^{14}$\,W/cm$^2$) infrared (IR) femtosecond pulse with a nonlinear medium (typically a noble gas). Under particular conditions, harmonic emission occurs every half optical cycle of the IR field (about 1\,fs), hence the XUV light is naturally emitted in the form of an attosecond pulse train (APT) \cite{Paul2001}. By gating the HHG process, it is possible to isolate single attosecond pulses (SAPs) from the train \cite{Drescher2002}. This sets the foundation for the development of a new branch of ultrafast spectroscopy: Attosecond Science. 
The low efficiency of the HHG process (typically of the order of $10^{-6}-10^{-7}$) limits the intensity of standard attosecond light sources, which are usually combined with a portion of the fundamental IR pulses in order to perform pump-probe spectroscopy without compromising the time resolution. First pioneering experiments based on time-resolved photoelectron spectroscopy have shown the capability of such a configuration to resolve ultrafast dynamics like Auger decay after inner core excitation in atoms \cite{Drescher2002}, tunnelling by strong field ionization \cite{Uiberacker2007}, electron motion in atoms \cite{Goulielmakis2010} and molecules \cite{Calegari2014} by using single attosecond pulses \cite{Sansone2010} or attosecond pulse trains \cite{Kelkensberg2011}. More recently, attosecond photoelectron spectroscopy has been used to study the ultrafast dynamics involved in the photoionization process \cite{Pazourek2015}, with particular attention to photoemission from solid targets \cite{Cavalieri2007,Locher2015,Neppl2015,Kasmi2017,Siek2017}. 
Complementary aspects of light-matter interaction can be studied with all-optical techniques. In this framework, attosecond transient absorption spectroscopy (ATAS) \cite{Gallmann2013,Geneaux2019} has established as a powerful tool for the investigation of sub-fs electron dynamics in insulators \cite{Schultze2013,Mashiko2016,Lucchini2016,Moulet2017}, semiconductors \cite{Schultze2014,Schlaepfer2018} and metals \cite{Volkov2019}. One of the key elements is the possibility to combine ATAS with a simultaneous, in-situ, calibration measurement of the IR pump electric field. This is normally done by injecting a noble gas as close as possible to the solid target and collecting the spectrum of the electrons photoemitted by the XUV radiation in the presence of the delayed IR pulse (see Sec. \ref{sec:pul}) while performing the ATAS measurement. In this way, it is possible to study the precise timing between the observed dynamics in the solid and the pump electric field \cite{Lucchini2020}, potentially revealing new physical phenomena \cite{Sato2018}.

\begin{figure*}[htbp]
	\centering
	\includegraphics[width=12cm]{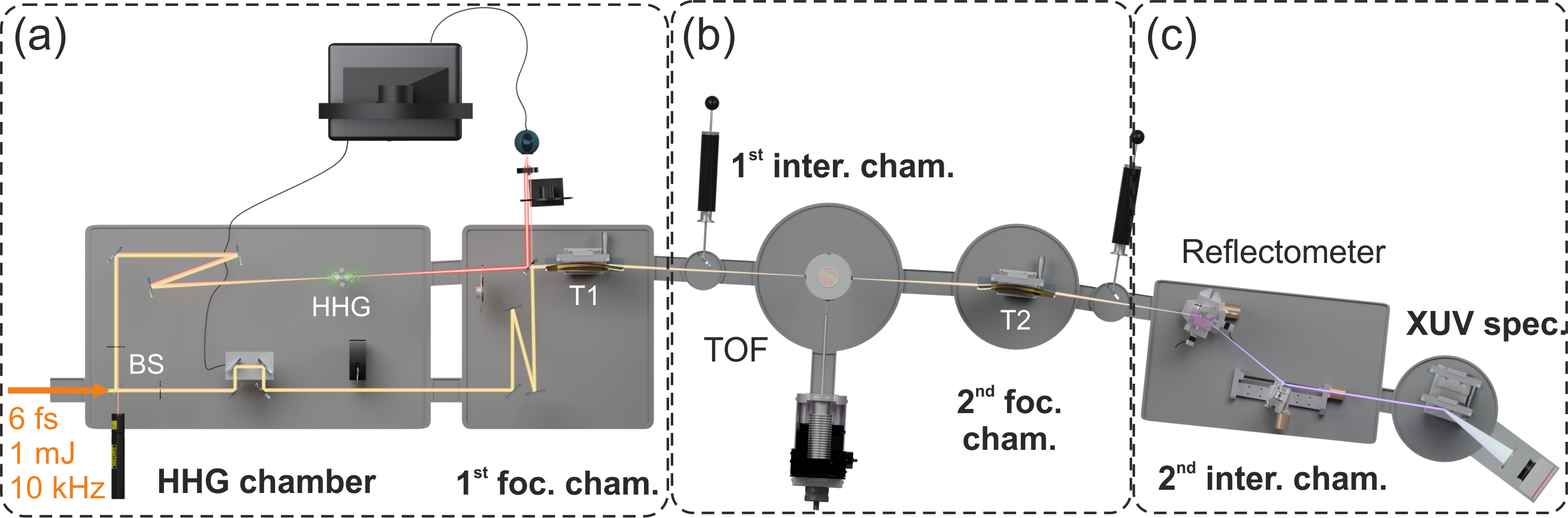}
	\caption{\label{fig:Full_setup} Orthogonal projection of the beamline setup. \textbf{(a)} HHG chamber and first focusing chamber which compose the actively stabilized interferometer. \textbf{(b)} First interaction chamber equipped with a TOF spectrometer, followed by the refocusing chamber. \textbf{(c)} Reflectometer for solid samples placed in the second interaction chamber together with an XUV spectrometer.}
\end{figure*}
Despite the outstanding results discussed above, the combination of high absorption coefficient in the XUV spectral range and relatively low photon flux of standard attosecond light sources, forces the samples to have thickness of the order of few tens of nanometers. Since this is not always possible, the applicability of the ATAS technique is strongly limited. Very recently, C.~J.~Kaplan and coworkers have shown that equivalent information can be obtained by looking at transient changes in the sample reflection rather than absorption \cite{Husek2017, Kaplan2018, Cirri2017, Mathias2012}.
Therefore, attosecond transient reflection spectroscopy (ATRS) candidates as a powerful tool to investigate ultrafast carrier dynamics in solids, which cannot be manufactured in thin membranes. While the applicability of ATRS in the few-fs domain has been already demonstrated \cite{Kaplan2019}, a clear observation of sub-fs dynamics is still missing. Furthermore, the more complex geometry needed to collect reflected rather than transmitted XUV radiation, makes it impossible to easily combine a reference gas target with ATRS. As a consequence, this hinders the possibility to access the attosecond timing of the observed dynamics with respect to the pump field. Here we report on the development and application of the first attosecond beamline for ATRS in a sequential double-foci geometry \cite{Locher2014}, which allows to overcome this limit. The beamline is composed by two separate interaction regions. The first region is equipped with a time-of-flight (TOF) spectrometer for photoelectron spectroscopy in gas targets. The second focus hosts a versatile XUV reflectometer. After a precise measurement of the phase delay induced by light propagation between the two foci\cite{Schlaepfer2017} (see Sec. \ref{sec:pph}), the beamline enables direct calibration of the pump-probe delay without introducing any additional geometrical limitation on the solid target, thus opening the possibility to investigate light-matter interaction in bulk samples with attosecond all-optical spectroscopy.

This paper is organized as follows: Section~\ref{sec:exp} presents the innovative beamline addressing separately each of its main constituents. In Sec.~\ref{sec:meas} we report on the results of some experimental measurements which validate our setup. 

\section{\label{sec:exp}Experimental setup}

The primary laser system is a commercial, 10-kHz laser system (AURORA, Amplitude Technologies), which generates pulses at a central wavelength of 800\,nm, with an energy of 2\,mJ, a full-width-half-maximum (FWHM) temporal duration of 25\,fs and stable carrier-envelope phase (CEP). Additional temporal pulse shortening is achieved through a hollow-core fiber compression system\cite{HCF} in combination with a battery of chirped mirrors. At the output of this stage, we obtain IR pulses with bandwidth ranging between 500 and 1200\,nm, a time duration of less than 6\,fs and pulse energy of about 1\,mJ. The compressed pulses enter the beamline (Fig.~\ref{fig:Full_setup}), composed of three main blocks: (\textit{a}) HHG chamber and actively stabilized interferometer, (\textit{b}) first interaction region for gas phase targets and (\textit{c}) second interaction region with an XUV reflectometer for solid samples.

\subsection{\label{sec:HHG}HHG and actively stabilized interferometer} 
\begin{figure*}[htbp]
    \centering
	\includegraphics[width=14cm]{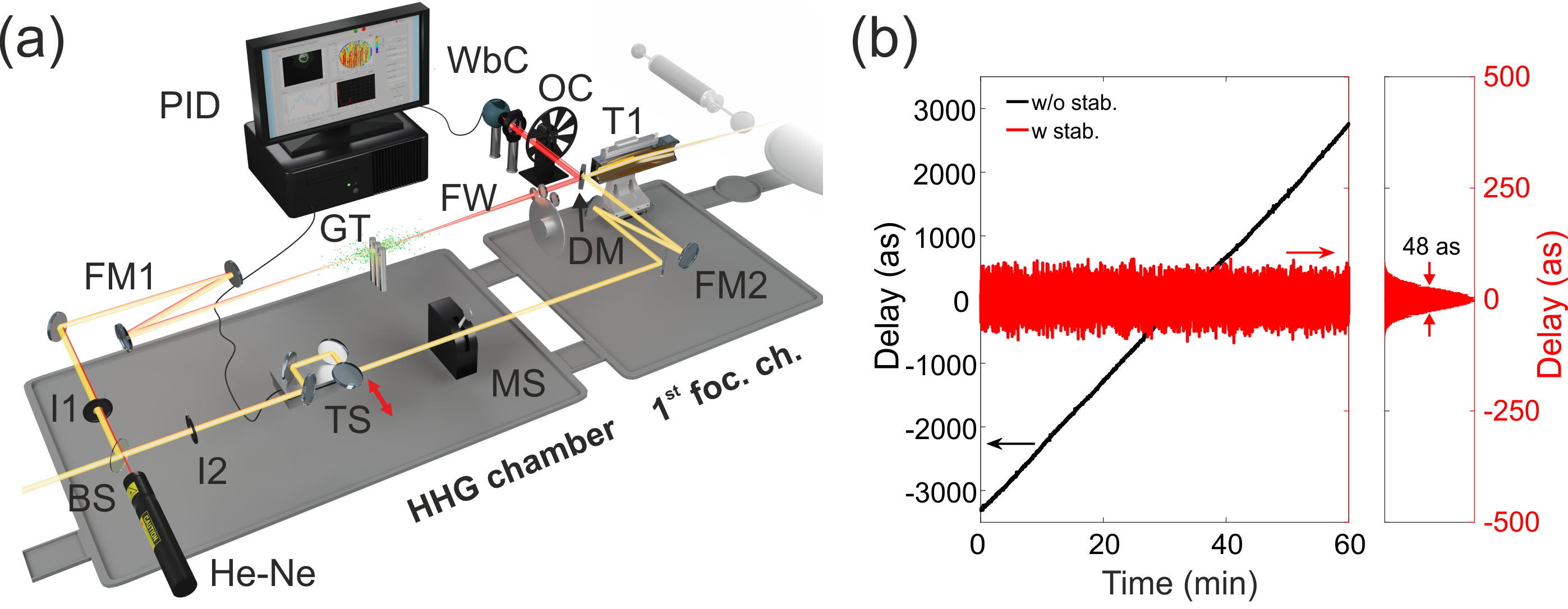}
	\caption{\label{fig:Interferometer}  \textbf{(a)} Schematic drawing of the HHG interferometer composed by two vacuum chambers (Fig. \ref{fig:Full_setup}(a)). The IR bean is represented in orange, the XUV in violet and the He-Ne used for active stabilization in red. The meaning of the labels which mark the optical elements is discussed in the text. \textbf{(b)} Comparison between the phase variation recorded by the webcam (WbC) with (red) and without (black) the active stabilization system, over a period of 60 minutes. The active stabilization system manages to correct the interferometer instabilities down to a standard deviation $\sigma$ of about 24\,as.}
\end{figure*}

Figure~\ref{fig:Interferometer}(a) shows the first main block of the beamline. After compression, the IR pulses enter a first evacuated chamber where they are split by a 70-30 broadband beam splitter (BS). The reflected part of the IR radiation (70$\%$) is focused by a converging mirror (FM1, $ROC= 1$\,m) onto a gas target (GT) for HHG. This latter is composed by three interchangeable 1-cm-thick gas cells, which are closed on both sides by an aluminum tape and filled with a noble gas (typically Ar). The IR laser drills an entrance and exit hole in the tape, providing efficient gas confinement (beam spot size on the target is $\sim 150$\,$\mathrm{\mu}$m). The cells are mounted on a motorized \textit{xyz}-translation stage allowing for fine alignment. This, in combination with a motorized iris (I1) placed on the beam path before HHG, allows to optimize the phase matching in real time. After XUV generation the beam enters a second evacuated chamber where the residual IR radiation is removed by a proper metallic filter. The filters are mounted on a filter wheel (FW) which enables different filter selection without breaking the vacuum. 

The transmitted part of the IR radiation (30$\%$) follows a delay line before being collinearly recombined with the harmonic radiation by a double-drilled mirror at 45$^\circ$ (DM), placed in the second chamber. The pulse energy is controlled through a motorized iris (I2), while the delay between the XUV and IR pulses is controlled by a retro-reflector mounted on a translation stage (TS) with nanometer resolution (Smaract SLC-2445-S-HV). A mechanical shutter (MS) placed on the delay line is used to achieve efficient noise subtraction in the measurement. The mechanical design of the shutter is based on the voice-coil actuator used in hard disk drivers\cite{Martinez2011}, which can be driven at a frequency of 1 Hz in vacuum for hours without overheating. A diverging mirror (FM2, $ROC= 1.2$\,m), placed in the second chamber before recombination, is used to match the IR beam divergence to the one of the XUV beam. In this way the two beams are focused onto the same plane by the first toriodal mirror (T1).

In order to perform time-resolved experiments with attosecond temporal resolution, very precise relative timing between pump and probe pulses is required. Any mechanical instability or thermal drift can introduce uncontrolled delay between the two interferometer arms, hence compromising the success of the experiment. Therefore, we implemented an active stabilization system. A frequency-stabilized helium-neon (He-Ne in Fig.~\ref{fig:Interferometer}) laser beam (at $\sim 633$\,nm) enters the interferometer through initial beamsplitter and follows the same paths of pump and probe pulses up to the double-drilled mirror. To avoid the He-Ne beam to be blocked by the metallic filters in the XUV path, we exploit the different propagation properties of this beam. Being characterized by a stronger spatial divergence, an annular part of the He-Ne beam passes through the glass support of the metallic filter \cite{Sabbar2014}. This portion of the beam is then reflected by the back of the double-drilled mirror and recombined with the He-Ne radiation which comes from the other arm of the interferometer and passes through the second hole in the mirror. The spatial interference fringes of these two beams are then measured by a web-cam (WbC), that is placed after an optical chopper (OC) suitably synchronized to remove the residual IR radiation. Using a digital PID controller acting on the delay stage, it is possible to stabilize the pump-probe relative delay with a residual standard deviation $\sigma = 24$\,as (Fig.~\ref{fig:Interferometer}(b)).

\subsection{\label{sec:ref}First interaction region and refocusing}
\begin{figure*}[htbp]
    \centering
    \includegraphics[width=14cm]{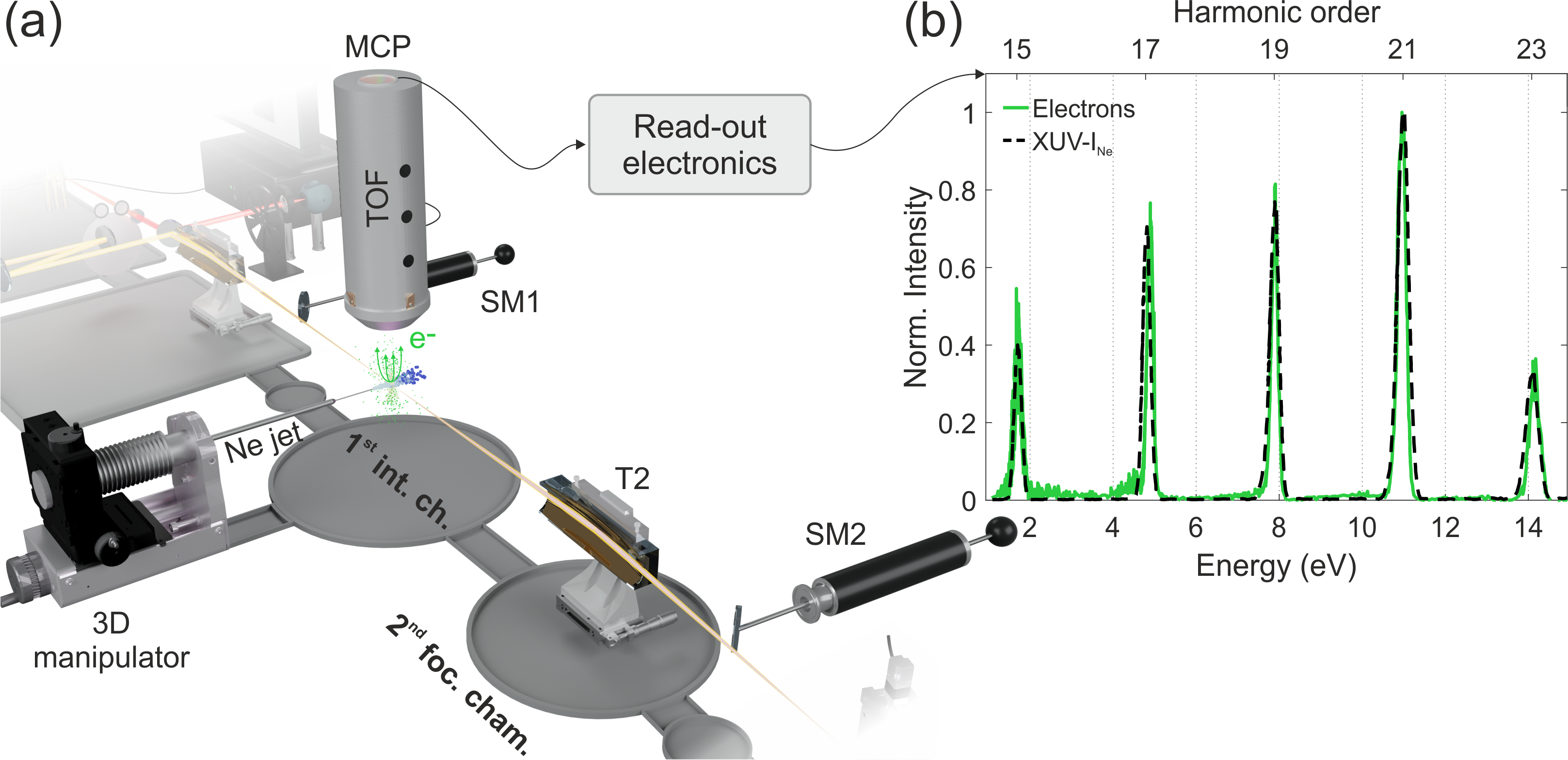}
	\caption{\label{fig:First_focus} \textbf{(a)} Schematic of the first interaction chamber and second refocusing mirror. A gas target is placed through an open needle in the first focus. The photoelectrons coming from the gas target are accelerated by static fields into a flying tube of a TOF before impinging on a micro-channel plate (MCP). The time evolution of the MCP current is recorded by a read-out electronics, leading to the photoelectron spectrum in \textbf{(b)} (green curve). The black dashed curve represents the ionizing XUV radiation as measuerd by the XUV spectrometer at the end on the line. The photon energy axis as been scaled by the gas ionization potential (in this case Ne, $I_{Ne} \simeq 21.56$\,eV).}
\end{figure*}
After recombination, XUV and IR pulses are focused by a gold-coated toroidal mirror (T1 in Fig.~\ref{fig:First_focus}(a)) with radii $R_{mer} = 14335$\,mm and $R_{sag} = 69.8$\,mm. Working at an incidence angle of $87^\circ$ we obtain one to one imaging with an arm length of 1000\,mm. A silver mirror (SM1) mounted on a push-pull manipulator can be inserted into the beam path soon after the first focusing chamber to extract the converging beams and control their spatial properties. Furthermore, removing the metallic filter from the XUV arm, it is possible to exploit spatial and spectral interferometry between the IR pump and the residual IR from the probe to set their temporal overlap.
The focus of the first toroidal mirror coincides with the center of a dedicated vacuum chamber equipped with an electron TOF spectrometer (eTOF v6 by Stefan Kaesdorf) which acquires the XUV photoelectron energy spectrum with a resolution of about 10\,meV. The target gas is injected in front of the TOF head with a stainless-steel needle (internal diameter of 500\,$\mathrm{\mu m}$), mounted on a 3D manipulator for fine spatial adjustment. Figure~\ref{fig:First_focus}(b) shows an example of photoelectron spectrum (green-solid curve) obtained by ionizing neon with a comb of harmonics (black-dashed curve). After proper correction for the transfer function of the TOF spectrometer and the Ne cross-section, electron and photon spectra can be nicely superimposed to one another. 

The TOF chamber is followed by a second gold-coated toroidal mirror (T2) with radii $R_{mer} = 11500$\,mm and $R_{sag} = 59$\,mm, incidence angle of $87^\circ$, focal length 800\,mm, which collects the radiation and makes a one-to-one image of the first focus onto the solid target placed in the reflectometer. A second silver mirror (SM2) mounted on a push-pull manipulator allows for beam extraction for the diagnostics of the second focus. 

\subsection{\label{sec:rfl}Second interaction region: XUV reflectometer} 

\begin{figure*}[htbp]
    \centering
    \includegraphics[width=14cm]{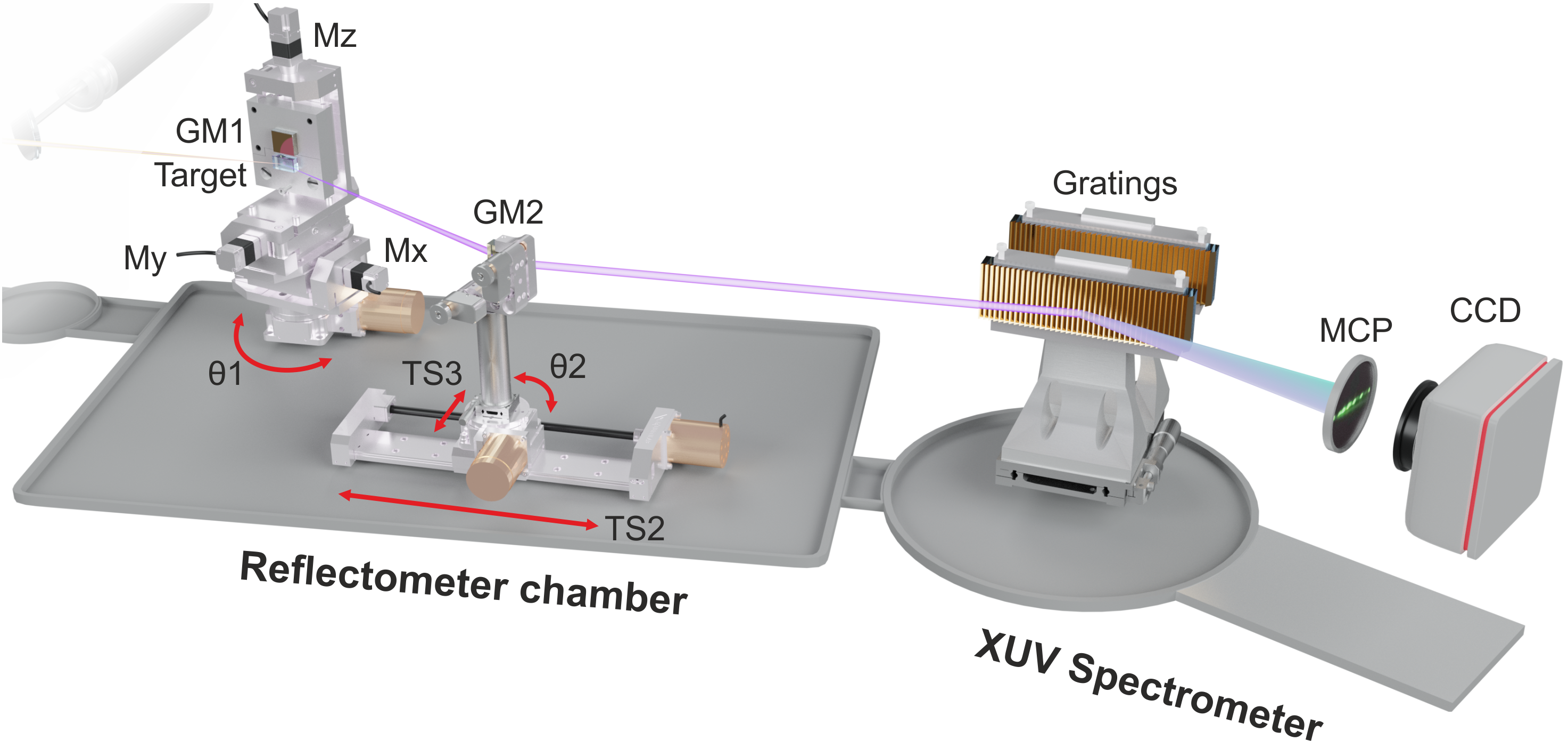}
	\caption{\label{fig:Reflectometer} Schematic of the XUV reflectometer. The solid sample is mounted on an holder that can be rotated along its vertical axis and translated in the three spatial directions. A reference gold mirror (GM1) is placed on top of the sample. A second gold mirror (GM2), mounted onto a roto-transnational stage, is used to collect the reflected radiation and steer it into the XUV spectrometer composed by a dispersive grating, an MCP and a phosphor screen. The image of the XUV spectrum which is formed on the phosphor is finally read by a CCD camera.}
\end{figure*}

The second focus coincides with the sample position of the XUV reflectometer (Fig.~\ref{fig:Reflectometer}). It consists of two main elements which allow to perform reflectivity measurements in a wide range of angles of incidence. The first element is a motorized sample holder which can translate along the $x$, $y$ and $z$ directions thanks to three linear motors (M$\mathrm{x}$, M$\mathrm{y}$, and M$\mathrm{z}$) with a resolution of 0.1\,$\mathrm{\mu m}$, and rotated around the vertical axis ($\mathrm{\theta_1}$) with a resolution of 0.01$^\circ$. The sample support is a steel plate with a 13-mm groove, designed to host an additional reference gold-coated mirror (GM1). This mirror can be moved in place of the sample to collect background or reference data during the experiment. The remaining degree of freedom, the vertical tilt, can be adjusted manually when a new sample is mounted for the first time in order to have the sample surface laying in the plane defined by the reference gold mirror surface. 

In order to improve the versatility of the XUV reflectometer, we intercept the reflected beam with a second gold-coated mirror (GM2) mounted onto a roto-translational stage. In this way, we can set the incident angle on the target in a range between 40$^\circ$ and 80$^\circ$ without the need for major rearrangement of the optics. The mirror is mounted on a motorized holder for fine horizontal and vertical tilt. Between the mirror holder and the rotator ($\theta_2$) we placed a manual translation stage (TS3) used to precisely align the rotation axis with the gold mirror surface. Finally, the whole support is mounted onto a long-range translator (TS2, model MP-21 by Micronix, with a resolution of 0.1\,$\mathrm{\mu m}$), able to shift the mirror by 200\,mm. Small variations of the angle of incidence onto the sample may produce a large change in the beam position alongside the propagation direction before the spectrometer. With this setup, we can precisely correct the beam position without the need to vent the chamber.

The XUV spectrometer is composed by a diffraction grating, which spatially disperses the attosecond radiation, and a space sensitive detector consisting in a micro-channel plate (MCP) followed by a phosphor screen\cite{Poletto2001}. Finally, a CCD camera records an image of the signal on the phosphor screen. The spectrometer chamber can host two different gratings, which can be moved into the beam path with an external manipulator. The first grating is an Hitachi model 001-0693 with central groove density of 600\,groves/mm, allowing for detection of photons in the range between 14\,eV and 62\,eV with a spectral dispersion of 18\,meV/pixel at 30\,eV (the pixel size is 30\,$\mathrm{\mu}$m on the phosphor screen). The second grating is an Hitachi model 001-0640 with a central groove density of 1200\,groves/mm and it is optimized for radiation in the spectral range between 27.5\,eV and 124\,eV, with a spectral dispersion of 11\,meV/pixel at 30\,eV. 

\section{\label{sec:meas}Experimental results}

In this section a few experimental measurements will be presented, which demonstrate the functionality and versatility of the beamline. In Sec.~\ref{sec:pul} we report on the generation and temporal characterization of SAPs. Section~\ref{sec:pph} describes a simultaneous double RABBITT (Reconstruction of Attosecond Burst By Two-photon Transitions)\cite{Paul2001,Muller2002}, which both demonstrates the high fidelity of the re-imaging and allows us to measure the phase difference between the two foci. Finally, Sec.~\ref{sec:rlf_m} contains an example of static and dynamical reflectivity measurements on an insulator (silicon dioxide) and a semiconductor (germanium).

\subsection{\label{sec:pul}Single attosecond pulse generation and characterization}

\begin{figure}[htbp]
	\centering
	\includegraphics[width=8.5cm]{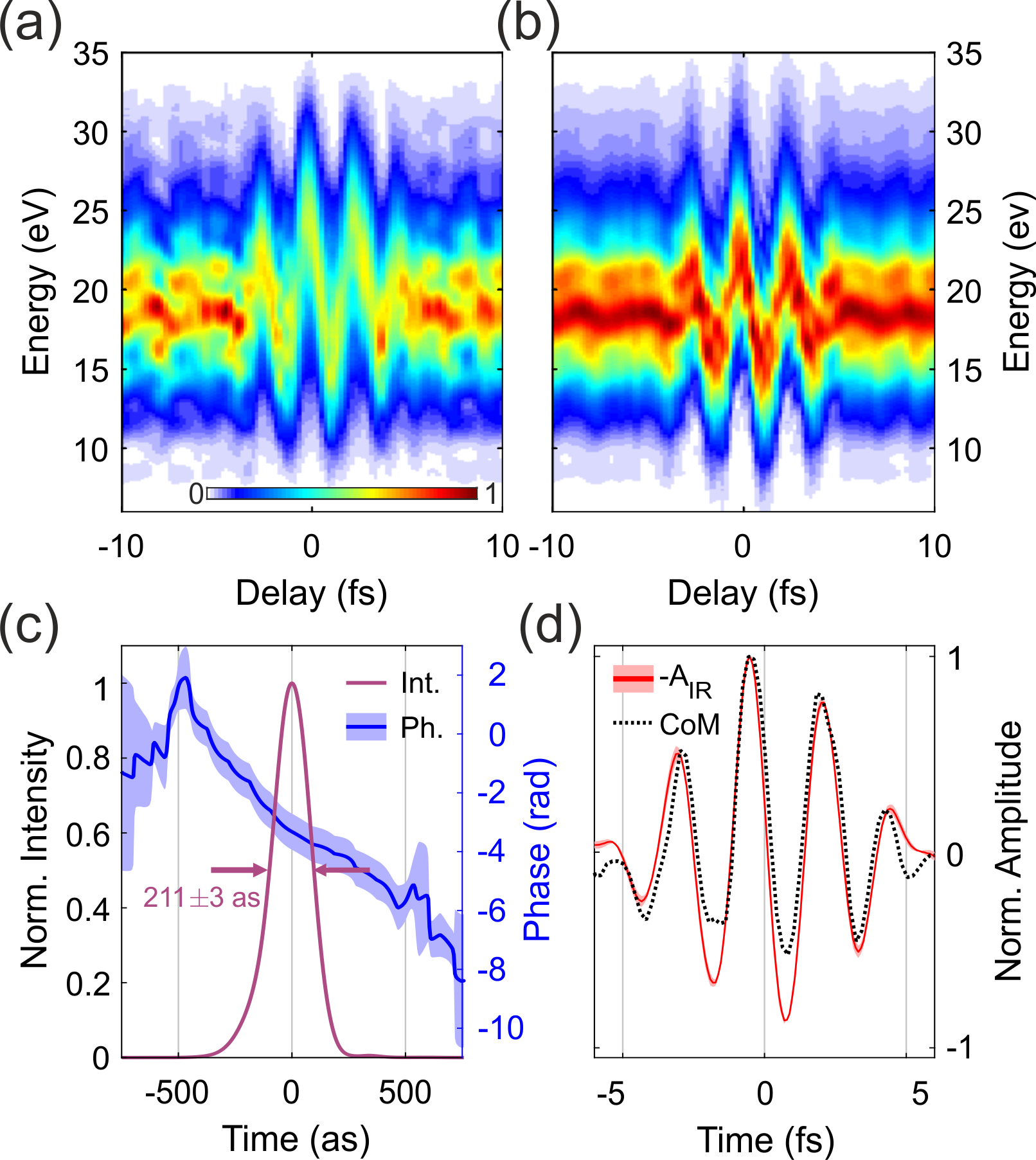}
	\caption{\label{fig:STK_rec} \textbf{(a)} Experimental streaking trace generated in Ne with an IR intensity of $(5.2 \pm 0.3)\times 10^{12}$\, W/cm$^2$. \textbf{(b)} Reconstructed trace after 2000 iterations of the ePIE algorithm. \textbf{(c)} Intensity (black curve) and phase (blue curve) temporal profiles of the SAP. \textbf{(d)} IR vector potential as reconstructed by the iterative algorithm (red curve) or as extracted from the center of mass of the trace in (a) (black-dotted curve). The solid lines in the reconstruction results of (c) and (d) represent the average over 20 reconstructions performed with different initial conditions. The shaded areas extend over twice the standard deviation.}
\end{figure}

We generated SAPs by employing the ionization gating technique \cite{Ferrari2010}. For sub-6-fs pulse duration, due to the relatively tight focusing, the driving IR pulses have enough intensity to ionized almost all the gas during their leading edge. When combined with CEP stabilization, this can lead to the direct generation of SAPs without the need for additional spectral or temporal manipulation. We characterized the pulses by performing a streaking measurement\cite{Itatani2002} in combination with FROG-CRAB (frequency resolved optical gating for the complete reconstruction of attosecond burst)\cite{Mairesse2005} reconstruction. We injected Ne gas in the first interaction chamber. The SAP photons are energetic enough to ionized the atoms and emit an electron in the continuum. The photoelectron spectrum is recorded by the TOF spectrometer. If a delayed IR pulse is added to the process, it behaves as an ultrafast phase modulator, changing the electron momentum. The collection of photoelectron spectra measured at different relative delays between the SAP and the IR pulse (Fig.~\ref{fig:STK_rec}(a)) is called spectrogram or streaking trace.
As the information it encodes is highly redundant, it is possible to retrieve both the IR and XUV temporal characteristics with an appropriate iterative reconstruction algorithm. Figure~\ref{fig:STK_rec}(b) shows the reconstructed trace after 2000 iterations of the extended ptychographic iterative engine (ePIE)\cite{Lucchini2015}, while the associated SAP and IR temporal characteristics are shown in Figs.~\ref{fig:STK_rec}(c) and \ref{fig:STK_rec}(d), respectively. In order to evaluate the uncertainty of our reconstruction procedure, we run the iterative code 20 times with different initial guesses for the pulses. The solid lines in Figs.~\ref{fig:STK_rec}(c) and \ref{fig:STK_rec}(d) represent the average value of the reconstruction outputs while the shaded area marks twice the standard deviation. The SAP has a FWHM time duration of $211\pm3$\,as and a residual quadratic chirp of -0.2981\,as$^2$. As expected, the reconstructed IR vector potential (red solid curve in Fig.~\ref{fig:STK_rec}) follows the center of mass of the spectrogram (black dashed curve in Fig.~\ref{fig:STK_rec}) and lasts for $4.7\pm0.3$\,fs (FWHM of the IR intensity envelope).

\subsection{\label{sec:pph}Simultaneous double RABBITT}

One of the main advantages of the sequential double-foci geometry adopted for this beamline is the possibility to run simultaneous measurements in the two interaction regions. In this way, one can perform an ATRS experiment while acquiring a spectrogram. The advantage of this approach is twofold. Not only it is possible to obtain a live characterization of the temporal properties of pump and probe pulses with the FROG-CRAB technique described above, but it is also possible to precisely calibrate the pump-probe delay axis with respect to the IR pulse field. In particular, the latter is necessary to access the attosecond timing of the observed dynamics and exploit the full potentials of the beamline. Since beam propagation between the two foci introduces an additional delay between the IR and XUV pulses, a precise calibration of this propagation delay (or equivalently, propagation phase) is necessary in order to obtain an absolute pump-probe delay axis. 
\begin{figure}[htbp]
	\centering
	\includegraphics[width=8.5cm]{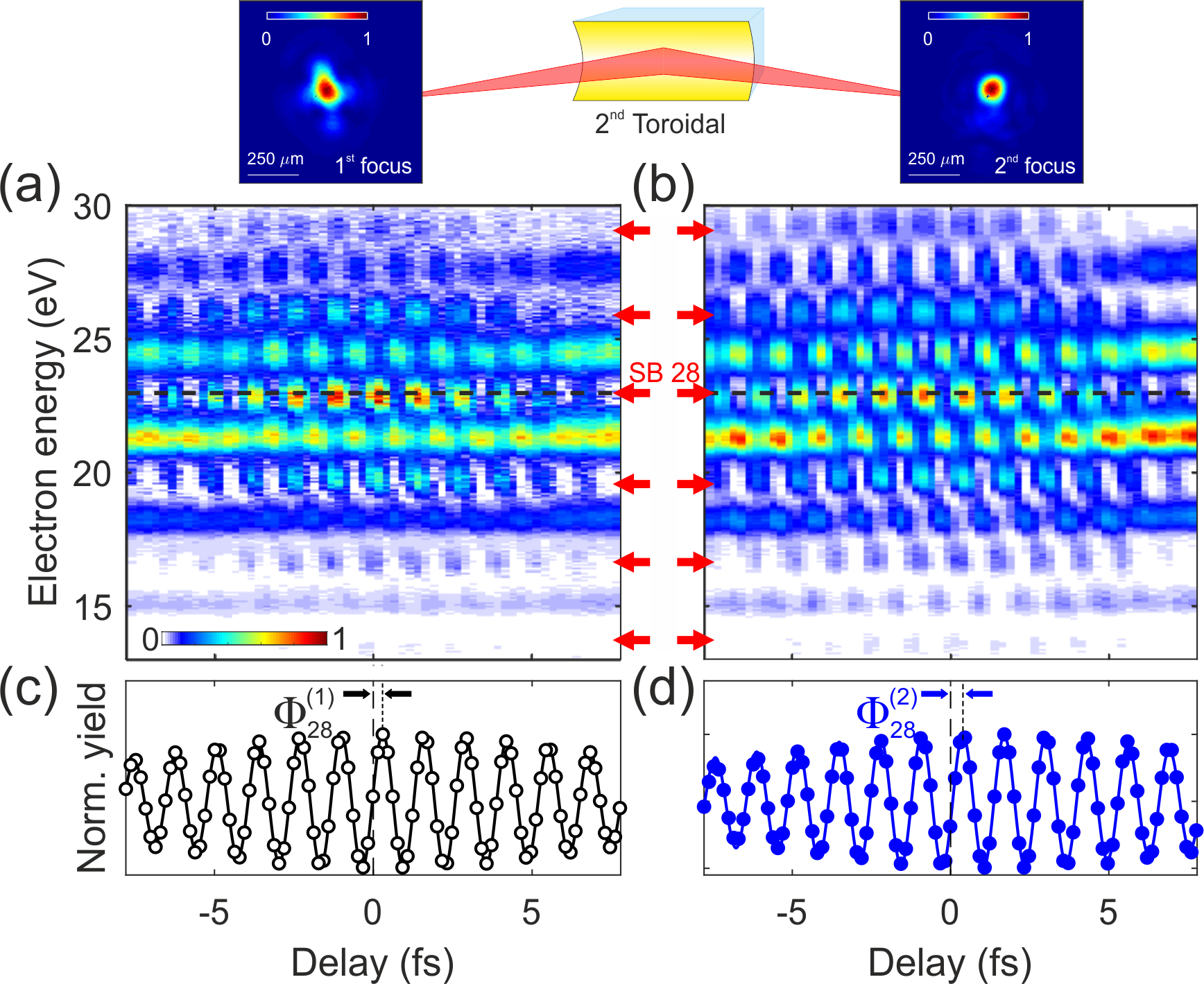}
	\caption{\label{fig:double_RAB} \textbf{(a)}, \textbf{(b)}, RABBITT traces acquired in the first and second focus, respectively. The signals appearing for small pump-probe delays and marked with the red arrows correspond to SB22 to SB32. \textbf{(c)}, \textbf{(d)}, SB28 yield obtained integrating the RABBITT trace in an energy band of 0.6\,eV around the black dashed horizontal line in (a) and (b), respectively. Open and close circles represent the data. The continuous line is the result of a fit with the function $SB(\tau)= a e^{-\left(\frac{\tau}{\sigma}\right)^2} \cos\left(\omega_{IR}\tau-\Phi_{SB}\right)$ where $\tau$ is the pump-probe delay, $\omega_{IR}$ is the IR field angular frequency and $\Phi_{SB}$ is the SB phase. The cartoon on top of the figure shows the IR beam profile measured in the two foci.}
\end{figure}

We measured the propagation phase with a double-RABBITT experiment. To this end, we placed a second TOF spectrometer, identical to the first one, in the second focal position. We generated an APT by using 25-fs IR pulses and used the harmonic comb to ionize two Ne targets and collect simultaneously two spectrograms (Figs.~\ref{fig:double_RAB}(a),~(b)). In the absence of the IR field, the photoelectron spectra are characterized by discrete peaks corresponding to the direct ionization by each odd harmonic. When the IR field is added, sideband (SB) peaks in between the main bands appear (red arrows in Figs.~\ref{fig:double_RAB}(a),~(b)). These peaks correspond to a two-color two-photon process, which involves the absorption of an XUV photon and the additional emission or absorption of an IR photon. As it originates by two energetically degenerate quantum paths, the SB intensity oscillates with the pump-probe delay at a frequency which is twice the IR frequency $\omega_{IR}$ (period of about 1.25\,fs). As an example, Figs.~\ref{fig:double_RAB}(c),~(d) show the electron yield associated to SB28 in the two foci. In the framework of the second order perturbation theory, it is possible to demonstrate that the harmonic phase is imprinted in the phase of the SB oscillation. Therefore, a direct comparison between the SB phases extracted in the first  and second focus measures the effect of pulse propagation as reported by F. Schlaepfer et al.\cite{Schlaepfer2017}.
\begin{figure}[htbp]
	\centering
	\includegraphics[width=8cm]{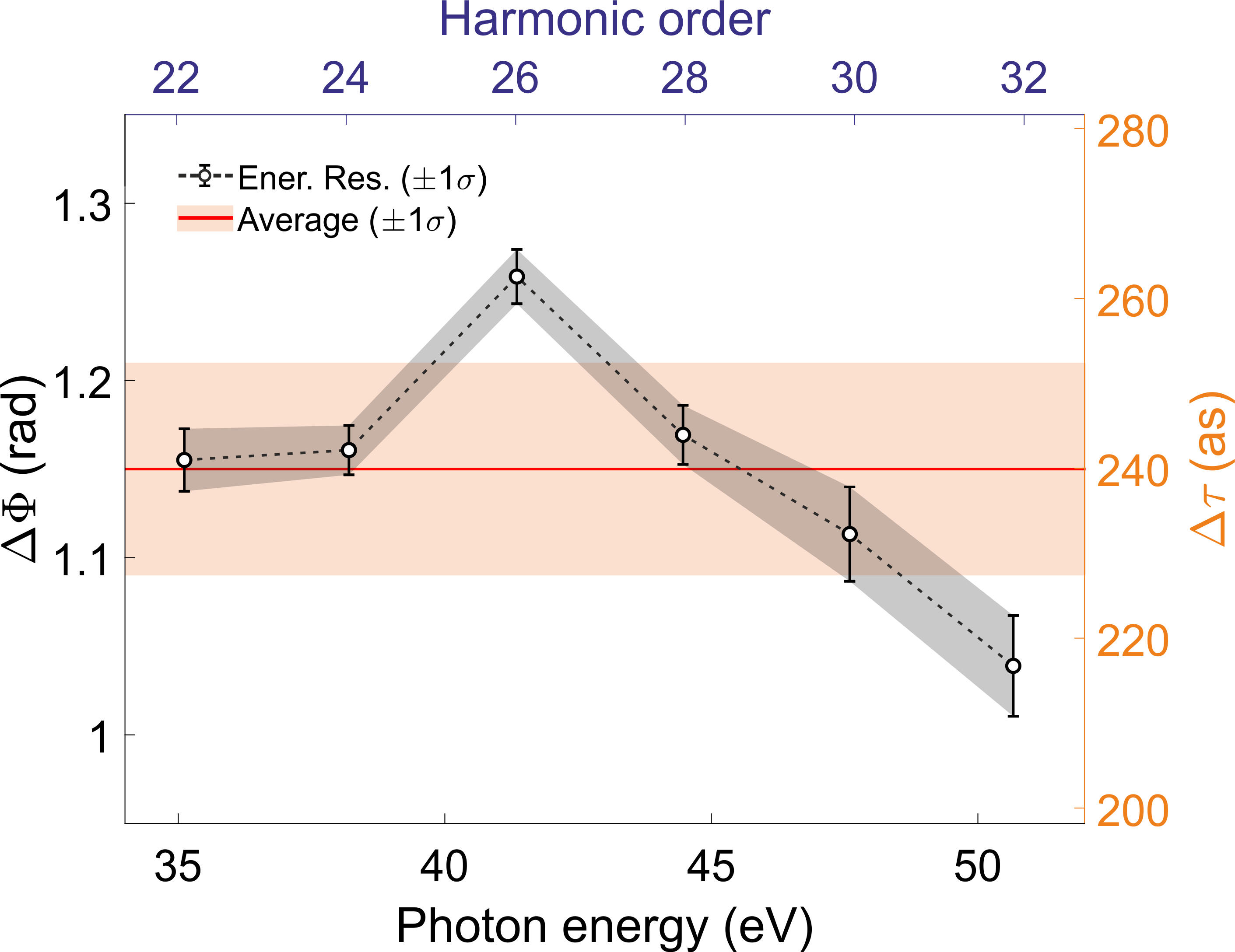}
	\caption{\label{fig:phase} Propagation phase/delay extracted from simultaneous double-RABBITT experiments as the one reported in Fig.~\ref{fig:double_RAB}. Since no particular dispersive optics are present in the beamline between the first and second focus, the propagation phase stays almost flat in the energy region under investigation (max variation of $\sim 25$\,as). Therefore, we can evaluate its effect by taking the average value of 240\,as~$\pm 13$\,as (red curve and shaded area).}
\end{figure}
In particular, the phase of the SB of order $2q$, $\Phi_{2q}$, can be decomposed in the sum of three terms: $\Phi_{2q} = \Delta\theta_{2q}+\Delta\theta_{at}-\theta_{IR}$. The first term originates from the phase difference between consecutive harmonics (it is the so called attochirp\cite{Varju2005}). The second term, $\Delta\theta_{at}$, is a target specific phase associated with the photoemission process. The last term accounts for the phase of the IR field. Since the two RABBITT traces are simultaneously recorded and with the same gas target (Ar), $\Delta\theta_{2q}$ and $\Delta\theta_{at}$ are the same in the two foci. Therefore, a non vanishing difference between $\Phi_{2q}^{(1)}$, measured in the first focus, and $\Phi_{2q}^{(2)}$, measured in the second focus, gives directly any additional phase which is imprinted onto the radiation upon propagation. Figure~\ref{fig:phase} shows the extracted phase difference $\Delta\Phi_{2q}=\Phi_{2q}^{(2)}-\Phi_{2q}^{(1)}$ from SB 22 to 32. Black open circles represent the average over 7 repeated measurements while the error bars extends over twice the standard deviation. Since $\Delta\Phi_{2q}$ originates from the difference in IR and XUV reflection phase on the second toroidal  mirror, and from a geometrical (Gouy phase) term, we expect it to be almost flat in the energy region of interest. Indeed our results show a maximum variation of $\Delta\Phi_{2q}$ of about 25\,as around its average value. Therefore, we can describe the propagation phase with a unique, energy-independent value given by the average of the different $\Delta\Phi_{2q}$ measured and amounting to $\Phi_{prop} = 1.15\pm 0.06$\,rad (red line and shaded area in Fig.~\ref{fig:phase}). Dividing by $2\omega_{IR}$, we obtain a propagation delay $\tau_{prop} = 240\pm 13$\,as, which can be used to calibrate the delay axis in the second interaction region.

\subsection{\label{sec:rlf_m}Reflectivity measurements}

Once the SAPs are generated and the beamline has been fully characterized, it is possible to perform ATRS with high temporal resolution. In order to prove the flexibility of our spectrometer and validate our calibration approach, we first measured the static reflectivity of a known target as amorphous silicon dioxide (SiO$_2$). 
\begin{figure}[htbp]
	\centering
	\includegraphics[width=8.5cm]{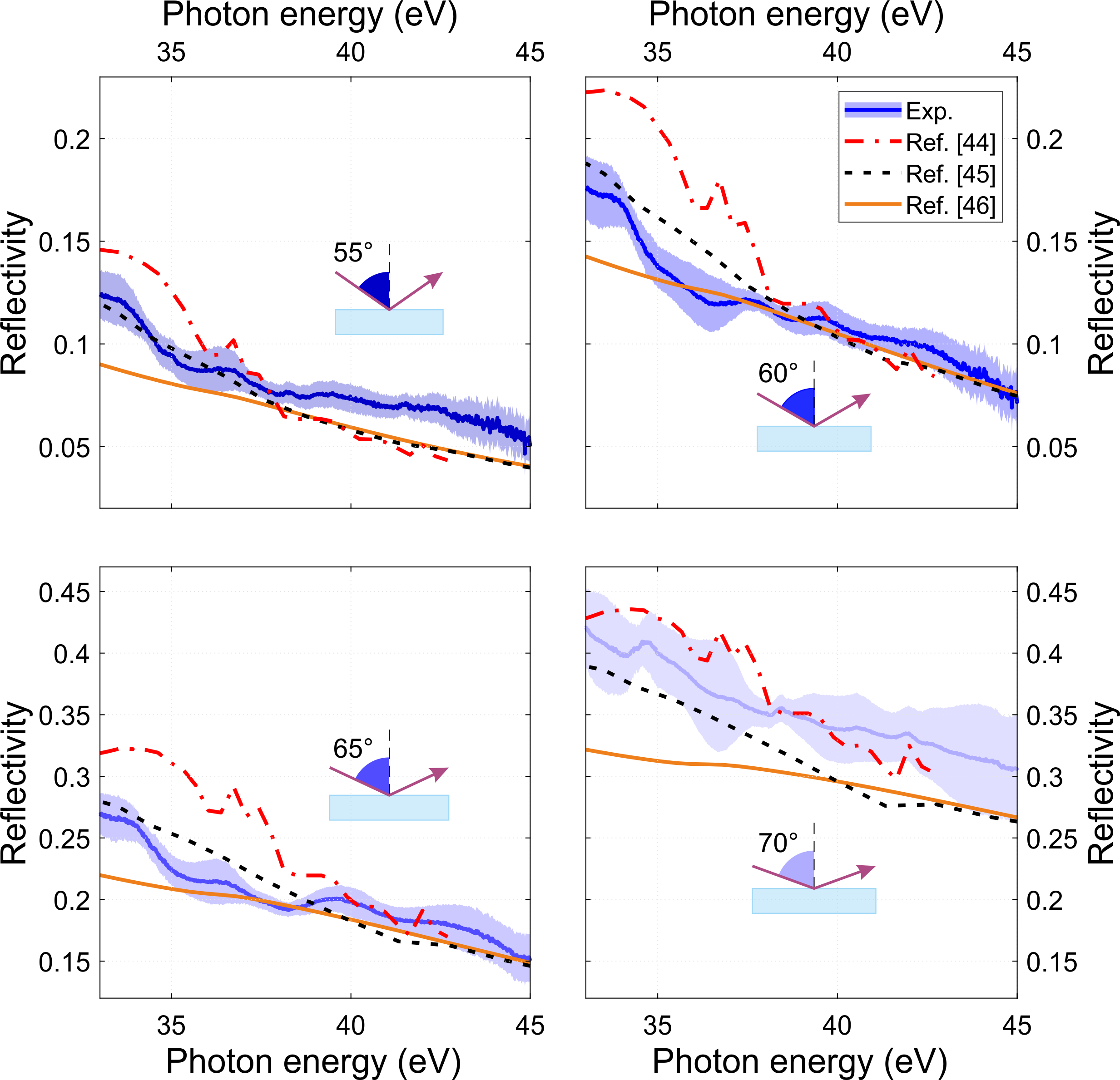}
	\caption{\label{fig:staticMeas}Measured static reflectivity of fused silica, $R_{SiO_2}$, at an incidence angle of: \textbf{(a)} $55^\circ$, \textbf{(b)} $60^\circ$, \textbf{(c)} $65^\circ$ and \textbf{(d)} $70^\circ$. In all panels the experimental reflectivity is marked by the blue solid curve (shaded area represented the standard deviation for repeated measurements), while red dash-dotted, black dashed and yellow solid curves display literature data as derived from the SiO$_2$ refractive index published in Refs. \cite{Tan2003}, \cite{palik1991handbook} and \cite{HENKE1993181}, respectively.}
\end{figure}
The absolute sample reflectivity $R_s$ is given by the ratio of the reflected, $I_s$, and the incident, $I_0$, XUV intensities. While $I_s$ is proportional to the signal measured by the XUV spectrometer at the end of the beamline, there is no direct way to measure $I_0$. Nevertheless, this problem can be solved by measuring the beam intensity, $I_{Au}$, reflected by a calibration target like a gold mirror. Since the gold reflectivity, $R_{Au} = I_{Au}/I_0$, is known from literature \cite{Werner2009}, the unknown sample reflectivity can be derived as follows:
\begin{equation}
    R_s = \frac{I_s}{I_0} = \frac{I_s}{I_{Au}}R_{Au} = \frac{S_s}{S_{Au}}R_{Au},  
\end{equation}
where $S_s$ and $S_{Au}$ are the XUV spectra measured after reflection onto the sample or the gold mirror, respectively. If the two spectra are measured under identical conditions, the procedure is justified.
\begin{figure*}[htbp]
	\centering
	\includegraphics[width=14cm]{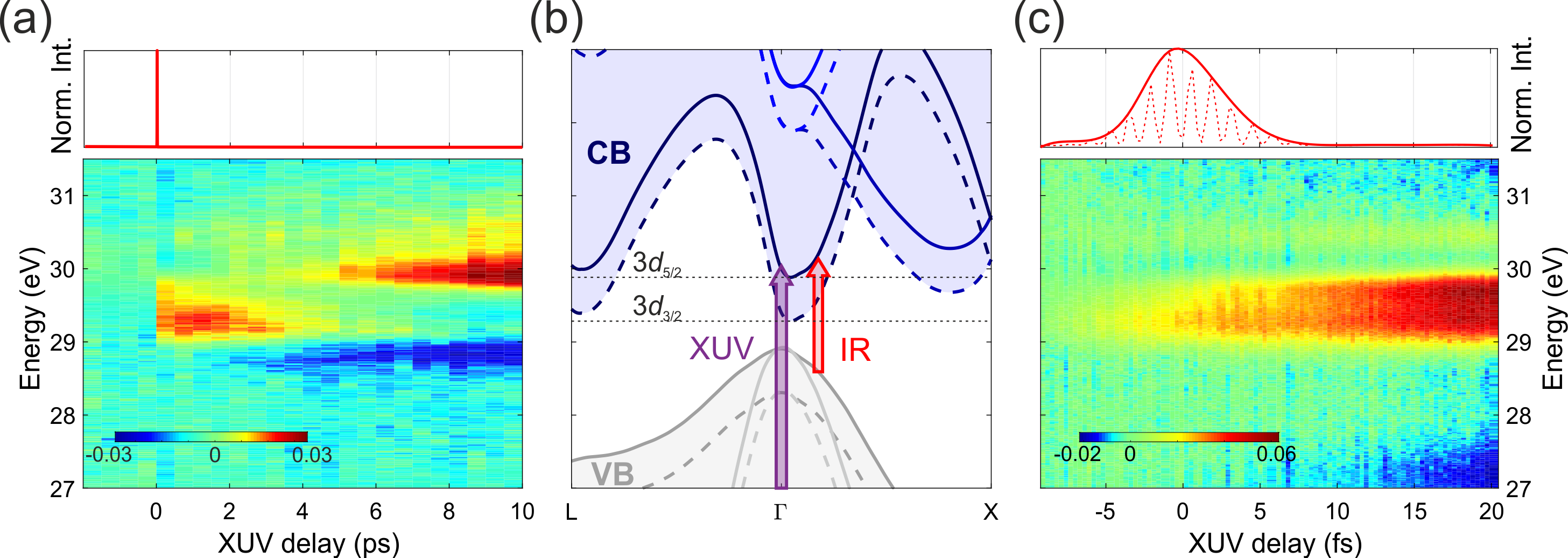}
	\caption{\label{fig:dynMeas}\textbf{(a)} Transient reflectivity $\Delta R/R$ measured from a Ge sample pumped with an IR intensity of $1.27\cdot 10^{13}$ W/cm$^2$ at an angle of incidence of 66$^\circ$. The top panel shows the square of the IR vector potential as retrieved from a simultaneous photoemission experiment. \textbf{(b)} Cartoon of the pump-probe mechanism. The IR photons promote electrons from the top of the valence band (VB) into the conduction band (CB). The dynamics thus initiated are then probed by the attosecond radiation via a transition from the Ge 3\textit{d} level to the VB and CB ($M_{4,5}$ edge). \textbf{(c)} Temporal zoom of (a). The upper panel in (a) and (c) report the IR intensity envelope as extracted from the simultaneous photoelectorn measurements. The red-dotted curve in (c) shows the square of the center of mass of the high energy part of the photoelectron spectrogram.}
\end{figure*}
Figure~\ref{fig:staticMeas} presents the static reflectivity of SiO$_2$, $R_{SiO_2}$, measured with the procedure described above for four incidence angles ranging from 55$^\circ$ to 70$^\circ$. The blue solid lines indicate $R_{SiO_2}$ obtained by averaging over 100 independent measurements, while the shaded area represents the measurement uncertainty (twice the standard deviation). In this case, as the sample is a very large fused silica plate (size of $25\times50$\,mm$^2$), we deposited a layer of 50 nm of gold onto a corner of the sample and used this gold layer as reference in order to assure high reproducibility of the measurements. The other solid curves in Fig.~\ref{fig:staticMeas} show $R_{SiO_2}$ as computed from literature data\cite{Tan2003,palik1991handbook,HENKE1993181}. Within the experimental error, we can conclude that our procedure for the measurement of energy-resolved absolute reflectivity is correct. 

As a further validation of the setup, we performed an ATRS experiment of a known sample. In particular, we measured the effect of a relatively strong IR pump in bulk germanium as reported by Kaplan et al. \cite{Kaplan2018}. We used a relatively short pulse train and recorded the IR-induced relative changes in the sample reflectivity defined as the difference between the Ge reflectivity with and without the IR pump, divided by the static Ge reflectivity without the pump: $\frac{\Delta R}{R}\big(E_{ph},\tau\big) = \nicefrac{\left[R^{IR}_{Ge}(E_{ph},\tau)-R^{wo}_{Ge}(E_{ph})\right]}{R^{wo}_{Ge}(E_{ph})}$. The results we obtained at an incidence angle of 66$^\circ$ (magical angle for Ge) and XUV photon energies ranging from 27 to 32\,eV, are reported in Fig.~\ref{fig:dynMeas}(a) together with the intensity envelope of the IR vector potential $A_{IR}(t)$ as extracted from the simultaneous photoemission measurement. Our findings fully reproduce what observed by Kaplan and co-authors. The IR field promotes electrons from the valence band (VB) to the conduction band (CB) initiating the dynamics, which are subsequently probed by the XUV pulse through the $M_{4,5}$ edge located at $\sim30$\,eV below the VB top (Fig.~\ref{fig:dynMeas}(b)). We observe a slow increase in reflectivity which develops on a picosecond time scale around 30.1 and 30.7\,eV, together with a slow decrease at 28-29\,eV. Between 29 and 30\,eV we find, instead, a prompt increase in reflectivity, which decays within 3\,ps. These features are associated to complex mechanisms of band-gap renormalization, hot electron and hole relaxation via inter-valley scattering and electron-phonon coupling via acoustic phonons. Since $\Delta R/R$ is sensitive to both the real and imaginary part of the dielectric function, a direct assignment of the observed features to these mechanisms is not trivial and requires a detailed analysis as reported in Ref.\cite{Kaplan2018}. 
Figure~\ref{fig:dynMeas}(c) shows a temporal zoom of $\Delta R/R$ in the region of pump-probe overlap. Since we used an APT to perform ATRS, the information on the IR CEP is lost and therefore it is not possible to extract the time-dependent IR amplitude from the simultaneous photoelectron spectrogram. Nevertheless, we can retrieve the IR pulse envelope (upper panel in Figs.~\ref{fig:dynMeas}(a),~(c)) which can be used to independently calibrate the pump-probe delay axis. We note the presence of some fast oscillations in the region of pump-probe overlap, between 29 and 30 eV. These oscillations are probably related to the pumping mechanism observed in Ref. \cite{Schlaepfer2018}, but their discussion is beyond the scope of this work.

These results prove that this setup is a reliable tool for assessing electron dynamics in solids with attosecond transient reflection measurements in a pump-probe configuration.

\section{\label{sec:sum}Summary}

We presented and validated a unique beamline for attosecond transient reflection spectroscopy in a sequential double-foci geometry. The attosecond streaking measurement reported in Sec.~\ref{sec:pul} demonstrate the capability of this setup to generate and characterize single attosecond pulses. Precise temporal calibration between the two interaction points was achieved by performing a simultaneous double-RABBITT experiment. The static absolute reflectivity of silicon dioxide reported in Sec.~\ref{sec:rlf_m} is in good agreement with what previously reported in literature, both showing the flexibility of this setup and confirming the experimental approach. Finally, the measured pump-induced relative changes in the reflectivity of germanium crystals demonstrate the possibility to perform simultaneous experiments in two different physical systems located in the two interaction regions. In particular, the possibility to perform a simultaneous calibration of the pump-probe delay with attosecond resolution fills the gap between transient absorption and reflection measurements, thus opening new perspectives for the investigation of ultrafast dynamics in solid with attosecond all-optical techniques.

\section*{\label{sec:ack}Acknowledgments}

The authors wish to thank Dr. Christian Rinaldi for providing the germanium samples.
This project has received funding from the European Research Council (ERC) under the European Union's Horizon 2020 research and innovation programme (AuDACE, grant agreement No. 848411). ML and GI further acknowledge fundings from MIUR PRIN aSTAR, Grant No. 2017RKWTMY. 

\section*{Data Availability}
The data that support the findings of this study are available from the corresponding author upon reasonable request.


\section*{References}
\bibliographystyle{Science}
\bibliography{biblio}

\begin{thebibliography}{10}

\bibitem{Krausz2009}
F.~Krausz, M.~Ivanov, {\it Reviews of Modern Physics\/} {\bf 81}, 163 (2009).

\bibitem{Nisoli2017}
M.~Nisoli, P.~Decleva, F.~Calegari, A.~Palacios, F.~Martín, {\it Chemical
  Reviews\/} {\bf 117}, 10760 (2017).

\bibitem{Paul2001}
P.~M. Paul, {\it et~al.\/}, {\it Science\/} {\bf 292}, 1689 (2001).

\bibitem{Drescher2002}
M.~Drescher, {\it et~al.\/}, {\it Nature\/} {\bf 419}, 803 (2002).

\bibitem{Uiberacker2007}
M.~Uiberacker, {\it et~al.\/}, {\it Nature\/} {\bf 446}, 627 (2007).

\bibitem{Goulielmakis2010}
E.~Goulielmakis, {\it et~al.\/}, {\it Nature\/} {\bf 466}, 739 (2010).

\bibitem{Calegari2014}
F.~Calegari, {\it et~al.\/}, {\it Science\/} {\bf 346}, 336 (2014).

\bibitem{Sansone2010}
G.~Sansone, {\it et~al.\/}, {\it Nature\/} {\bf 465}, 763 (2010).

\bibitem{Kelkensberg2011}
F.~Kelkensberg, {\it et~al.\/}, {\it Phys. Rev. Lett.\/} {\bf 107}, 043002
  (2011).

\bibitem{Pazourek2015}
R.~Pazourek, S.~Nagele, J.~Burgd\"orfer, {\it Rev. Mod. Phys.\/} {\bf 87}, 765
  (2015).

\bibitem{Cavalieri2007}
A.~L. Cavalieri, {\it et~al.\/}, {\it Nature\/} {\bf 449}, 1029 (2007).

\bibitem{Locher2015}
R.~Locher, {\it et~al.\/}, {\it Optica\/} {\bf 2}, 405 (2015).

\bibitem{Neppl2015}
S.~Neppl, {\it et~al.\/}, {\it Nature\/} {\bf 517}, 342 (2015).

\bibitem{Kasmi2017}
L.~Kasmi, {\it et~al.\/}, {\it Optica\/} {\bf 4}, 1492 (2017).

\bibitem{Siek2017}
F.~Siek, {\it et~al.\/}, {\it Science\/} {\bf 357}, 1274 (2017).

\bibitem{Gallmann2013}
L.~Gallmann, {\it et~al.\/}, {\it Molecular Physics\/} {\bf 111}, 2243 (2013).

\bibitem{Geneaux2019}
R.~Geneaux, H.~J. Marroux, A.~Guggenmos, D.~M. Neumark, S.~R. Leone, {\it
  Philosophical Transactions of the Royal Society A: Mathematical, Physical and
  Engineering Sciences\/} {\bf 377} (2019).

\bibitem{Schultze2013}
M.~Schultze, {\it et~al.\/}, {\it Nature\/} {\bf 493}, 75 (2013).

\bibitem{Mashiko2016}
H.~Mashiko, K.~Oguri, T.~Yamaguchi, A.~Suda, H.~Gotoh, {\it Nature Physics\/}
  {\bf 12}, 741 (2016).

\bibitem{Lucchini2016}
M.~Lucchini, {\it et~al.\/}, {\it Science\/} {\bf 353}, 916 (2016).

\bibitem{Moulet2017}
A.~Moulet, {\it et~al.\/}, {\it Science\/} {\bf 357}, 1134 (2017).

\bibitem{Schultze2014}
M.~Schultze, {\it et~al.\/}, {\it Science (New York, N.Y.)\/} {\bf 346}, 1
  (2014).

\bibitem{Schlaepfer2018}
F.~Schlaepfer, {\it et~al.\/}, {\it Nature Physics\/} {\bf 14}, 560 (2018).

\bibitem{Volkov2019}
M.~Volkov, {\it et~al.\/}, {\it Nature Physics\/} {\bf 15}, 1145 (2019).

\bibitem{Lucchini2020}
M.~Lucchini, {\it et~al.\/}, {\it Journal of Physics: Photonics\/}  (2020).

\bibitem{Sato2018}
S.~A. Sato, {\it et~al.\/}, {\it Physical Review B\/} {\bf 98}, 1 (2018).

\bibitem{Husek2017}
J.~Husek, A.~Cirri, S.~Biswas, L.~{Robert Baker}, {\it Chemical Science\/} {\bf
  8}, 8170 (2017).

\bibitem{Kaplan2018}
C.~J. Kaplan, {\it et~al.\/}, {\it Physical Review B\/} {\bf 97}, 1 (2018).

\bibitem{Cirri2017}
A.~Cirri, J.~Husek, S.~Biswas, L.~R. Baker, {\it Journal of Physical Chemistry
  C\/} {\bf 121}, 15861 (2017).

\bibitem{Mathias2012}
S.~Mathias, {\it et~al.\/}, {\it Proceedings of the National Academy of
  Sciences\/} {\bf 109}, 4792 (2012).

\bibitem{Kaplan2019}
C.~J. Kaplan, {\it et~al.\/}, {\it J. Opt. Soc. Am. B\/} {\bf 36}, 1716 (2019).

\bibitem{Locher2014}
R.~Locher, {\it et~al.\/}, {\it Review of Scientific Instruments\/} {\bf 85},
  013113 (2014).

\bibitem{Schlaepfer2017}
F.~Schlaepfer, {\it et~al.\/}, {\it Optics Express\/} {\bf 25}, 3646 (2017).

\bibitem{HCF}
M.~Nisoli, S.~De~Silvestri, O.~Svelto, {\it Applied Physics Letters\/} {\bf
  68}, 2793 (1996).

\bibitem{Martinez2011}
S.~Mart\'inez, {\it et~al.\/}, {\it Review of Scientific Instruments\/} {\bf
  82}, 046102 (2011).

\bibitem{Sabbar2014}
M.~Sabbar, {\it et~al.\/}, {\it Review of Scientific Instruments\/} {\bf 85},
  103113 (2014).

\bibitem{Poletto2001}
L.~Poletto, G.~Tondello, P.~Villoresi, {\it Review of Scientific Instruments\/}
  {\bf 72}, 2868 (2001).

\bibitem{Muller2002}
H.~G. Muller, {\it Applied Physics B: Lasers and Optics\/} {\bf 74}, 17 (2002).

\bibitem{Ferrari2010}
F.~Ferrari, {\it et~al.\/}, {\it Nature Photonics\/} {\bf 4}, 875 (2010).

\bibitem{Itatani2002}
J.~Itatani, {\it et~al.\/}, {\it Physical Review Letters\/} {\bf 88}, 173903
  (2002).

\bibitem{Mairesse2005}
Y.~Mairesse, F.~Qu{\'{e}}r{\'{e}}, {\it Physical Review A\/} {\bf 71}, 011401
  (2005).

\bibitem{Lucchini2015}
M.~Lucchini, {\it et~al.\/}, {\it Optics Express\/} {\bf 23}, 29502 (2015).

\bibitem{Varju2005}
K.~Varj\'u, {\it et~al.\/}, {\it Journal of Modern Optics\/} {\bf 52}, 379
  (2005).

\bibitem{Tan2003}
G.~L. Tan, M.~F. Lemon, R.~H. French, {\it Journal of the American Ceramic
  Society\/} {\bf 86}, 1885 (2003).

\bibitem{palik1991handbook}
E.~Palik, G.~Ghosh, {\it Handbook of Optical Constants of Solids\/}, Academic
  Press handbook series (Elsevier Science, 1991).

\bibitem{HENKE1993181}
B.~Henke, E.~Gullikson, J.~Davis, {\it Atomic Data and Nuclear Data Tables\/}
  {\bf 54}, 181  (1993).

\bibitem{Werner2009}
W.~S. Werner, K.~Glantschnig, C.~Ambrosch-Draxl, {\it Journal of Physical and
  Chemical Reference Data\/} {\bf 38}, 1013 (2009).

\end{thebibliography}


\end{document}